**High sensitivity nanoporous gold metamaterials for plasmonic sensing**

*Denis Garoli, Eugenio Calandrini, Giorgia Giovannini, Aliax Hubarevich, Vincenzo Caligiuri and Francesco De Angelis*

Dr. D. Garoli, Dr. E. Calandrini, G. Giovannini, A. Hubarevich, V. Caligiuri, F. De Angelis
Istituto Italiano di Tecnologia – Via Morego, 30, I-16163 Genova, Italy
E-mail: francesco.deangelis@iit.it

Surface Plasmon Resonance sensors are a well-established class of sensors which includes a very large variety of materials and detection schemes. However, the development of portable devices is still challenging as due to the intrinsic complexity of the optical excitation/detection schemes. In this work we show that Nanoporous gold (NPG) films may overcome these limitations by providing excellent sensitivity without the need of sophisticated fabrication approaches and/or optical setup. The sensing mechanism is related the co-localization of optical energy and analytes into the pores that promote an enhanced light-matter coupling. As result, when molecules are adsorbed into the pores, the NPG film shows a significant spectral shift of the effective plasma frequency and then an abrupt change of the reflectivity. By monitoring the reflectivity in the spectral region close to the plasma frequency (namely the plasma edge) is then possible to detect the analyte. Through a set of experiments we demonstrated a sensitivity exceeding 15.000 nm/RIU in the Near Infrared Range that is comparable with the state of the art of plasmonic metamaterials.

## Introduction

Biosensing technologies are emerging for fast, real-time identification of biomarkers and to quantify several species of interest at the same time, i.e. multiplex detection [1][2]. Most of these requirements can be satisfied leveraging on optical confinement and enhancement provided by plasmonic resonances. Surface plasmons resonances (SPR) are extremely sensitive to the dielectric permittivity of the environment in contact to the metal surface[3,4]. Hence, when molecules are adsorbed on plasmonic nanostructures the spectral positions of their resonances experience significant shifts even for few molecule binding events. The sensitivity of these class of sensors is therefore usually expressed as $S = \Delta\lambda/\Delta n$, i.e. the ration between the spectral shift $\Delta\lambda$ of the plasmon resonance and the change of the refractive index $\Delta n$ surrounding the sensor which are both induced by the presence of the analyte. Along the years, a wide range of optical excitation and detection schemes have been proposed togheter with the use of several kinds of nanostructures [5][6]. Among them, two main classes can be indentified: i) those based on propating plasmons, usually exctited via grating coupling, that allows the implementation of a microarray format for multiplexed and high-throughput biosensing platform [7][8]; ii) those based on localized resonances in individual nanostructures of different sizes and shapes, either dispersed in colloidal solution or fabricated on substrates by lithographic methods [9][10][11][12]. Such a long standing challenge recently culminated in plasmonic metamaterial based sensors showing a sensitivity approaching 30.000 nm/RIU[13][14] in the near infrared region. However, high sensitivity is very often associated with expensive fabrication methods or sophysticated detection schemes. In this respect Nanoporous Gold (NPG) offers a valid alternative by providing low cost fabrication methods and simple excitation schemes of strong plasmonic resonances[15]. Among the several newly investigated plasmonic materials[16], NPG presents significant advantages mainly due to the large surface/volume ratio. This latter not only ensures more binding sites not achievable with bulk metals[17][18][19][20] but, in a previous work we also demonstrated that NPG behaves like a metamaterial whose effective dielectric response can to be tailored by changing pore size and porosity, or equivalently by changing the fractal dimension[21]. In particular, the plasma frequency can be shifted over a wide spectral range of infrared wavelengths. Importantly, the porous metamaterial exhibits superior plasmonic properties compared to its bulk counterpart.[21–30] These properties, combined with a higher skin depth in the order of 100–200 nm, enables the penetration of light deep into the nanopores where the analyte can be accumulated. Such an efficient co-localization of analytes and optical energy promotes a robust light-matter coupling and high detection sensitivity[23].

In this work, we show that the spectral postion of effective plasma frequency (or the plasma edge) of nanoporous plasmonic metamaterials is extremely sensitive to the index of refraction of the surrounding environment. Hence it can be used as high performance SPR sensor. Through a set of different experiments we showed a sensity exceeding 18.000 nm/RIU at $\lambda$=1550 nm. This result can be achieved with very easy excitation/detection scheme that may consist in a reflectance measurent at fixed angle and wavelenght. It does not require complex optical setups and/or adavanced nanostructuring thus opening the route to portable SPR nanosensors with extreme sensitivity and low cost.

**Results and Discussion**

The preparation of NPG films followed well established procedures (previously reported in details[21]) in which an optimized dealloying process of 3 hours was carried out. Fabrication parameters were adjusted in order to have the plasma frequency in the NIR band. The concept for plasma frequency or plasma edge sensing is illustrated in Figure 1 together with a representative SEM image of the fabricated samples. The plasma edge is the spectral band, near the plasma frequency, in which the material switches from reflective to transparent[21]. As result of the enhanced light-matter coupling occurring in nanopores, the plasma edge of NPG is extremely sensitive to the index of refraction of the material filling the nanopores. By measuring the reflectivity is then possible to measure the plasma edge shifts and then to monitor the amount of analyte present into the nanopores.

In this first work we focused on the correct estimation of the sensitivity. The latter is not trivial because the very large surface area provided by nanoporous materials can bring to misleading interpretation. We approached this problem in three different ways which gave comparable results.

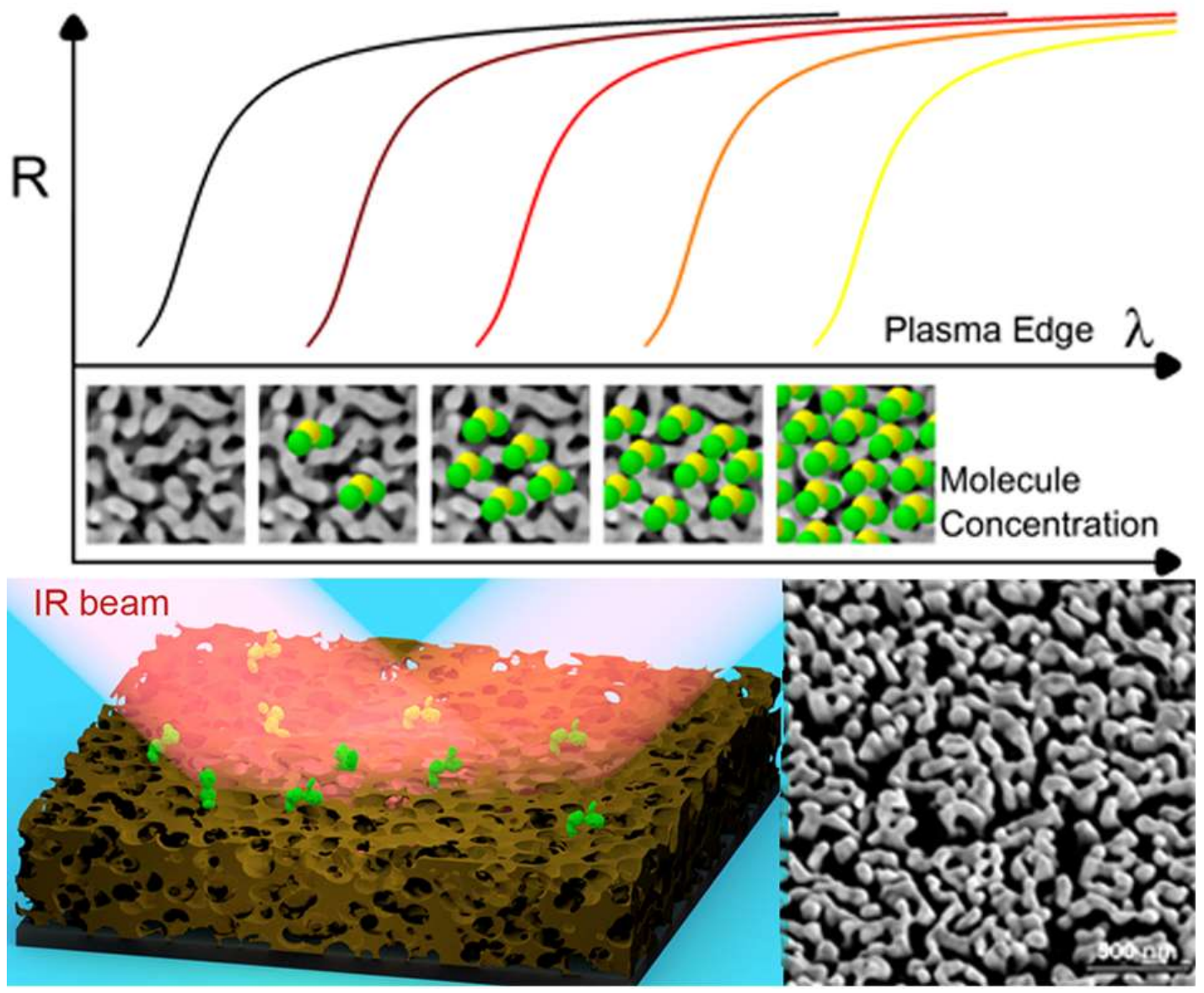

***Figure 1***. *Sensing approach – a Near Infrared light beam illuminates the NPG and the reflectance is measured around the plasma edge region where a significant shift can be detected.*

For first, we monitored the plasma edge shifts of NPG after successive depositions of very thin layers of $SiO_2$ by Atomic Layer Deposition (ALD) that allows a conformal coating of uniform thickness all over the NPG surface. Importantly it enables an exact control of the amount of material deposited that is why we chose this method. [31,32] After each deposition of 1.5 nm in thickness of $SiO_2$ (final thickness 9 nm) we measured the reflectance via Fourier Transform InfraRed (FTIR) spectroscopy in the spectral range spanning from 1500 to 3000 nm.

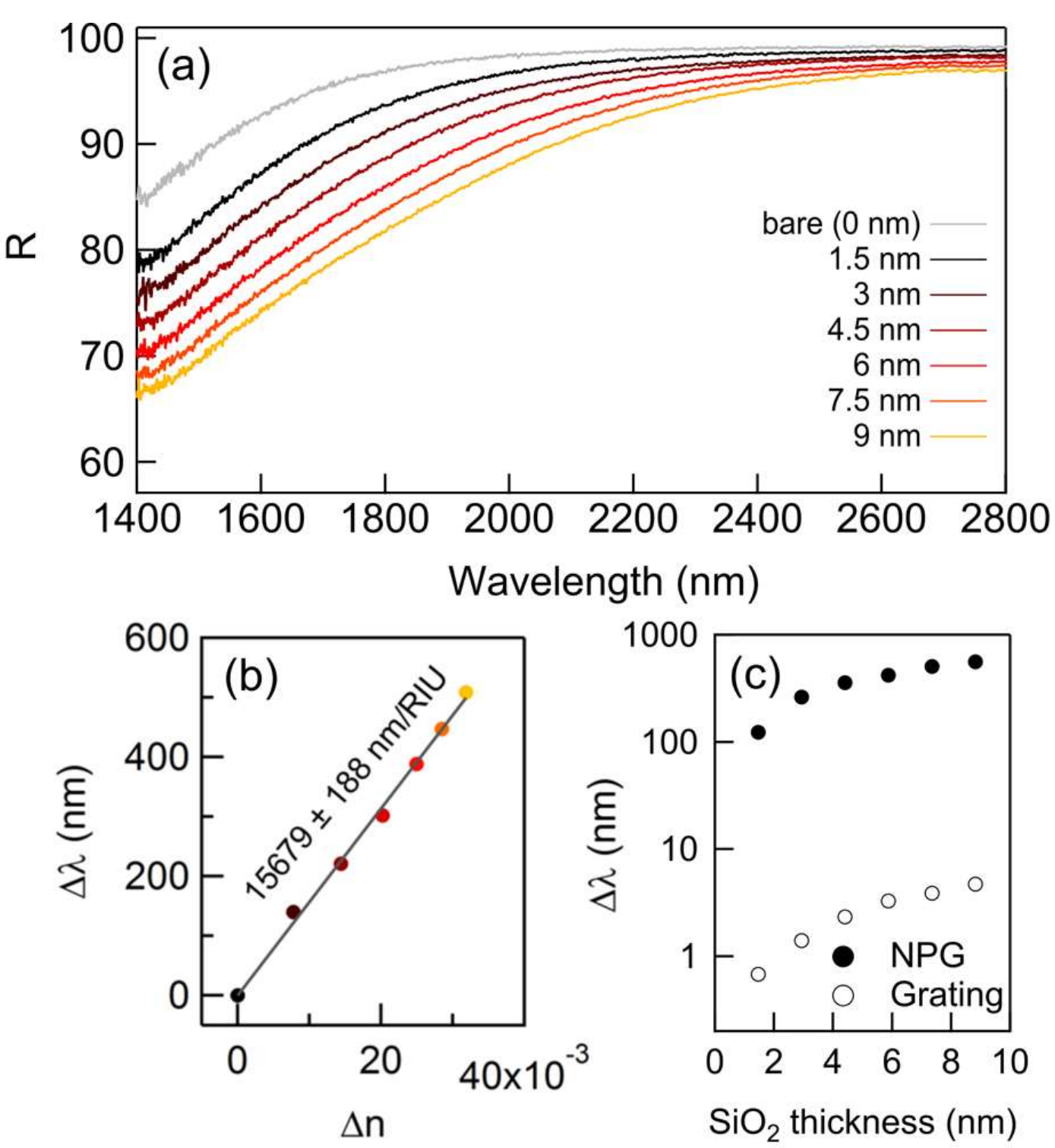

***Figure 2***. *NPG sensing performance with thin dielectric coating deposition. (a) Reflectance curves after successive $SiO_2$ layers depositions; (b) Spectral shift measured at R(0.85%); (c) Calculated sensitivity using the formula (1); (d-e) Comparison with the spectral shift obtained with the same dielectric coatings on a standard bulk gold grating* [33]

Results are reported in Figure 2: while increasing the thickness of the $SiO_2$ layer, reflectivity significantly drops and plasma edge redshifts even for vanishingly small quantities of $SiO_2$. We verified that the observed phenomena is a real plasma edge redshift, we prepared a sample also comprising 3D nanoantennas following a procedure recently reported. The increasing thickness of $SiO_2$ layer also shifts the SPR dip related to the antenna, i.e. both the plasma edge and the antenna resonances are extremely sensible to the environment.

The spectral shifts extracted from figure 2a are reported in figure 2b. For comparison we computed the shifts achievable with a conventional approach such as grating coupled SPR [33](GC-SPR) working in the same spectral range. As it can be seen, NPG exhibits spectral shifts which are 100 times larger than those of a GC-SPR under the same conditions. In agreement with well established literature, our simulations estimate a GC-SPR sensitivity of about 1500 nm/RIU[33]. This would suggest an empiric sensitivity of 1500×100=150.000 nm/RIU for NPG sensor. However, this value must be corrected by a factor due to the higher surface area of NPG with respect to bulk grating. Even though an accurate measurement of this parameter is not simple, a factor between 5 and 10 has been reported in several papers[34]. Hence, a conservative estimation of the sensitivity of NPG sensor is in the order of 15.000 nm/RIU that is comparable with the state of art of metamaterials[13,14,35].

As second approach to evaluate the sensitivity consists in treating the NPG-$SiO_2$ material as Insulator-Metal multilayered metamaterial, whose overall effective dielectric function is dominated by the component longitudinal to film plane, which can be easily calculated applying the interface conditions for the electric field vector (seem Methods for details).[36,37] By using this approximation we can calculate the refractive index shift $\Delta n(\omega,t)$ induced by the depostions of $SiO_2$. Given $\Delta n(\omega,t)$, the sensitivity expressed in nm/RIU follows from the well-known figure of merit:

$$S(\omega,t) = \frac{\lambda_t(R) - \lambda_0(R)}{\Delta n(\omega,t)} \qquad (1)$$

Where $\lambda_t$(R)- $\lambda_0$(R) is the spectral shift of the reflectivity at a fixed value *R* as a function of the thickness variation $\Delta t$ of deposited $SiO_2$. Results reported in Figure 2b show an average sensitivity of 15.700 nm/RIU in good agreement with the empiric evaluation achieved by comparing NPG with a conventional grating.

The third assessment of the sensitivity has been performed measuring the spectral shifts of NPG plasma edge in liquid phase by using solution of glycerol in water. This procedure has been previously reported in several papers [13][35] and it can be considered a standard method. Since water presents significant absorbance lines in the NIR band, we exploited a dedicated microfluidic chamber[38] that enabled to achieve reproducible measurements. The NPG film has been integrated in a 10 microns thick chamber with an IR transparent $CaF_2$ top window. Before each experiment NPG was treated in $O_2$ plasma (100W, 100% $O_2$ for 60 seconds) in order to guarantee the wettability of the porous matrix. Another critical aspect observed during this "in-liquid" measurements is the washing step after every data collection for a particular concentration of glycerol. We observed that the organic molecules tend to deposit on the NPG surface after some minutes. This create a not-uniform layer of "dirty" metal with not-reproducible optical response (see SI for an example). The microfluidic system developed for this experiment has been optimized in order to be extendable, i.e. it can be open and closed to access the substrate for cleaning or washing procedure. In this way the NPG film can be washed in ethanol after each measurement. The washing step has been verified by checking the NPG surface in SEM and by measuring the optical response in $H_2O$ at the end of the experiment.

Figure 3 illustrates the shift of the plasma edge obtained for glycerol concentrations between 1 and 10% m/V, which are typically used in literature [13]. The corresponding variations of index of refraction $\Delta n$ are equal to 0.00113, 0.00227, 0.005764, and 0.01179, respectively[39]. These measurements allow to directly evaluate the sensitivity by measuring the spectral shift of the reflectivity $\Delta\lambda(R)$ for different values of the index of refraction of the environment (water+glycerol). Figures 3b and 3c report the measured shifts and the obtained sensitivity that resulted in good agreement with the evaluations discussed above.

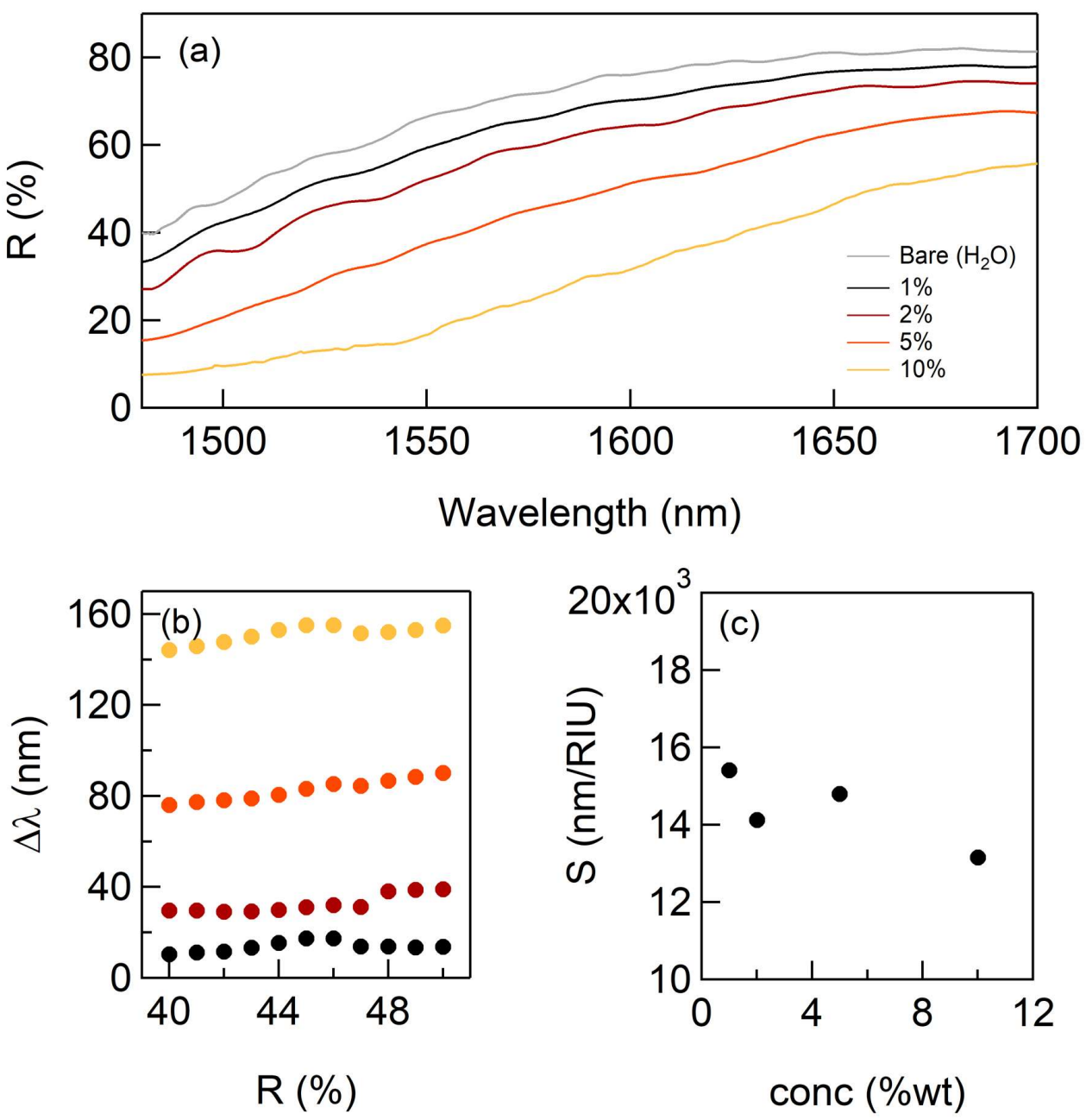

**Figure 3**. NPG sensing performance in $H_2O$+Glycerol. (a) Reflectance curves with different concentrations of glycerol; (b) Spectral shift measured at different R (between 40 and 50%); (c) Obtained sensitivity in term of $\Delta\lambda/\Delta n$.

As final test, we validated the performance of the proposed sensor by detecting low molecular weight proteins in solution. Liquid samples containing different concentration of polyhystideine – 7HIS (c.a. 1 kDa) were analyzed previous functionalization of the sensor surface. Here we mention that by using this approach a detection limit of 10 pM was achieved that is in agreement with what obtained by using state of the art metamaterial sensors [13][40].

The reported sensitivity can be explained by considering the plasmonic response of NPG in the NIR range. As illustrated in our recent experiments, the field enhancement and confinement provided by the NPG at the metal/air or metal/liquid interface is much higher than that provided by its bulk counterpart. The porous structure enables a spontaneous co-localization of the plasmonic field and the analyte thus promoting an effective interaction between them.

We remark that similar results can be achieved without the need of measuring the spectral response. In fact, after a proper sensor calibration, a direct measurement of the reflectivity at fixed wavelength and fixed angle can give an estimation of the analyte concentration. In perspective, such a scheme would lead to the development of cheap and portable SPR sensors.

## Conclusions

We demonstrated that nanoporous gold films show an effective plasma frequency or plasma edge whose spectral position is extremely sensitive to the index of refraction of the material filling the nanopores. Therefore, NPG films can be used as very sensitive SPR sensors. Both the optical excitation and detection schemes are very easy and do not require sophisticated fabrications or experimental setups. A sensitivity exceeding 15.000 nm/RIU was demonstrated by using different approaches. In light of these features we think these class of metamaterials may lead to a novel generation of portable SPR sensors.

## Methods

### *Optical spectroscopy*

The optical response in the IR range is investigated by means of FT-IR reflectance microscopy measurements by using a Nicolet™ iS™ 50 FT-IR Spectrometer coupled to the Nicolet Continuµm Infrared Microscope by Thermo Scientific™. A gold mirror was used as the reference. The optical response of this effective porous conductor was investigated employing a Drude-Lorentz model. Within this context, its dielectric permittivity is characterized by a reduction of the absolute values with respect of the bulk counterpart, that mitigates the optical losses and increases the skin depth enabling the penetration of the electromagnetic field deeply into the porous matrix loaded with the analyte to be detected.

In order to assess the benefits of NPG in a SPR configuration the far-field optical response of the fabricated film has been measured by means of FTIR spectroscopy using a microscope with a 30°-incidence Cassegrain objective in the reflection mode.

*Refractive index shift calculation*

In the case of $SiO_2$ deposition, the refractive index variation Δn is empirically calculated under the following assumption: the film covered by the $SiO_2$, can be tough as Insulator-Metal multilayered metamaterial, whose overall effective dielectric function is dominated by the component longitudinal to film plane, which can be easily calculated applying the interface conditions for the electric field vector. The dielectric functions used for these estimations are found in literature for gold and silicon dioxide, while for NPG is used the dielectric function previously calculated. Finally, recalling that $n = \sqrt{e}$, the following relation can be established

$$\Delta n = \sqrt{\varepsilon_{npg}} - \sqrt{\varepsilon_{avg//}} = \sqrt{\varepsilon_{npg}} - \sqrt{\frac{\varepsilon_{npg} t_{npg} + \varepsilon_{SiO2} t_{SiO2}}{t_{npg} + t_{SiO2}}}$$

Where $\varepsilon_{npg,SiO2}$ and $t_{npg,SiO2}$ are the dielectric functions and the thicknesses of the NPG and the silicon dioxide respectively. Under the above assumption, the refractive index shifts are reported in SI (Supporting Fig. S7) as a function of the wavelength for a 200nm thick NPG film.

*Surface functionalization protocol*

The surface was incubated for two hours at 40°C with 5mM (5mL) solution of Succinic anhydride in water. The surface was washed with water (5mL) to remove the unreacted anhydride and the carboxyl groups now exposed at the surface were activated with 5mM EDC 5mM NHS solution for 30 minutes at R.T. The coupling reagents were removed by flushing the surface with water and afterwards the surface was treated with a solution of polyhydtidine (concentration between 10 pM up to 10 nM) at R.T. After 2 hours the surface was washed once and analysed under both wet (in water) and dry condition.

**Acknowledgements**

The research leading to these results has received funding from the European Research Council under the Horizon 2020 Program, FET-Open: PROSEQO, Grant Agreement n. [687089].

**Competing financial interests:**
The authors declare no competing financial interests.